\documentclass[aps,prl,twocolumn,superscriptaddress]{revtex4}
\usepackage{graphicx,color,ulem}

\begin{document}

\title{Destruction of N$\rm \acute{e}$el order and appearance of superconductivity in electron-doped
cuprates by oxygen annealing process}

\author{Shiliang Li}
\affiliation{
Department of Physics and Astronomy, The University of Tennessee, Knoxville, Tennessee 37996-1200, USA
}
\author{Songxue Chi}
\affiliation{
Department of Physics and Astronomy, The University of Tennessee, Knoxville, Tennessee 37996-1200, USA
}
\author{Jun Zhao}
\affiliation{
Department of Physics and Astronomy, The University of Tennessee, Knoxville, Tennessee 37996-1200, USA
}
\author{H.-H. Wen}
\affiliation{
National Laboratory for Superconductivity, Institute of Physics and National Laboratory for Condensed Matter Physics,
Chinese Academy of Sciences, P. O. Box 603, Beijing 100080, P. R. China
}
\author{M. B. Stone}
\affiliation{
Neutron Scattering Sciences Division, Oak Ridge National Laboratory, Oak Ridge, Tennessee 37831-6393, USA
}
\author{J. W. Lynn}
\affiliation{
Center for Neutron Research, National Institute of Standards and Technology,Gaithersburg,
MD 20899-6102, USA
}
\author{Pengcheng Dai}
\email{daip@ornl.gov}
\affiliation{
Department of Physics and Astronomy, The University of Tennessee, Knoxville, Tennessee 37996-1200, USA
}
\affiliation{
Neutron Scattering Sciences Division, Oak Ridge National Laboratory, Oak Ridge, Tennessee 37831-6393, USA
}

\begin{abstract}
We use thermodynamic and neutron scattering measurements to 
study the effect of oxygen annealing on the superconductivity and magnetism in Pr$_{0.88}$LaCe$_{0.12}$CuO$_{4-\delta}$.
Although the transition temperature $T_c$ measured by susceptibility
and superconducting coherence length
increase smoothly with gradual
oxygen removal from the annealing process, bulk superconductivity, marked by a specific heat anomaly at $T_c$ and the presence of a neutron magnetic resonance, only appears abruptly when $T_c$ is close to the largest value.
These results suggest that the effect of oxygen annealing must be first determined in order to establish a Ce-doping dependence of antiferromagnetism and superconductivity phase diagram for electron-doped copper oxides.
\end{abstract}


\maketitle

\section{INTRODUCTION}
High transition temperature (high-$T_c$) superconductivity occurs in copper oxides when sufficient holes or electrons are doped into the CuO$_2$ planes of their antiferromagnetic (AF) parent compounds. To understand the relationship between antiferromagnetism and superconductivity, it is essential to determine the doping evolution of the magnetic phase diagram. For hole-doped  La$_{2-x}$Sr$_x$CuO$_4$ (LSCO), the static long-range AF order is suppressed and superconductivity emerges for a critical Sr-doping level of $x>0.06$ \cite{kastner,wakimoto}. In the case of electron-doped copper oxides such as Nd$_{2-x}$Ce$_x$CuO$_{4-\delta}$ (NCCO), Pr$_{2-x}$Ce$_x$CuO$_{4-\delta}$ (PCCO), and Pr$_{1-x}$LaCe$_x$CuO$_{4-\delta}$ (PLCCO),
the long-range AF phase appears to extend over a wide Ce-doping range ($x\geq 0.1$) and coexist with superconductivity  \cite{uefuji,fujita}. In particular, transport \cite{dagan,yu} and optical \cite{zimmer} measurements on PCCO thin films revealed an AF quantum critical point (QCP) inside the superconducting dome at $x\approx 0.16$, where the coexisting AF and superconducting phase ($0.13\leq x\leq 0.16$) is separated from the pure superconducting phase ($x>0.16$). These results are consistent with
the presence of a normal-state gap (induced by the static/fluctuating AF order) at $x=0.15$ on NCCO revealed by
 angle resolved photoemission spectroscopy \cite{matsui} and optical measurements \cite{onose}.
However, a recent neutron scattering experiment on NCCO suggests that
genuine long-range AF order and superconductivity do not coexist, and the QCP separating the
pure AF and superconducting phases
lies just before the superconducting dome at $x\approx 0.13$ \cite{motoyama}. Therefore,
the true AF phase in NCCO extends only to $x\approx 0.13$ and the phase diagrams of electron-doped materials are much different from previous work \cite{uefuji,fujita,dagan,yu,zimmer}, as illustrated in the inset of Fig. 1(a).

Unlike their hole-doped counterparts, electron-doped materials exhibit no superconductivity in the as-grown state until they are annealed in an oxygen-poor
environment to remove a small amount of oxygen \cite{tokura,takagi}.
Although three-dimensional (3D) static long-range AF order is observed to coexist with superconductivity in annealed NCCO for $0.13\leq x\leq 0.17$ \cite{uefuji,fujita}, the existence of static AF order has been argued as an artifact induced by the inhomogeneous annealing process for this family of materials \cite{motoyama}. Therefore, to understand the microscopic mechanism of the
annealing process and its impact on bulk superconductivity \cite{higgins,gauthier,kang}, one must choose electron-doped materials that can be annealed
continuously into pure superconductors without coexisting static AF order \cite{kim,jiang}. Our prior elastic neutron scattering measurements showed that electron-doped PLCCO ($x=0.12$) is ideal for this purpose \cite{wilson2}.

In this paper, we present systematic specific heat and neutron scattering
measurements designed to study the annealing effect on PLCCO ($x=0.12$) as the system is tuned from an as-grown antiferromagnet to a pure superconductor ($T_c=27$ K) without static AF order. Although magnetic susceptibility and elastic neutron scattering measurements suggest a gradual increasing $T_c$ (Fig. 1) and decreasing $T_N$ respectively with increasing oxygen loss \cite{wilson2}, we find that the bulk superconductivity and neutron magnetic resonance \cite{wilson} occur only for PLCCO samples with the largest $T_c$'s. Our results thus suggest that the neutron scattering results determined using NCCO single crystals without systematically studying the oxygen annealing effect is unreliable \cite{motoyama}, and the phase diagram obtained on PCCO films with properly tuned optimal superconductivity \cite{dagan,yu} reveals the intrinsic properties of the electron-doped materials.

\section{EXPERIMENT}
We grew PLCCO ($x=0.12$) single crystals by the traveling-solvent floating-zone method \cite{wilson2}.
The as-grown nonsuperconducting PLCCO exhibits static AF order with $T_N\sim 200$ K \cite{wilson2}. The annealing process is a dynamic process of removing oxygen, and the final oxygen content is determined by the temperature, the oxygen partial pressure, and the annealing time \cite{kim,jiang}. Instead of annealing in argon gas \cite{fujita}, we use
high vacuum annealing to reduce the oxygen partial pressure and increase the dynamic range of temperature \cite{kim}.
Figure 1(a) shows $T_c$ vs. oxygen loss $\Delta$ obtained for one PLCCO crystal ($\sim 1$ gram) using a thermogravimetric
analyzer (TGA) at temperatures from 800 to 880$^\circ$C. Data points correspond to sequential days of annealing.
Magnetic susceptibility measurements in Fig. 1(b)
show that the superconducting volume fraction becomes larger than 90\% for samples with $T_c\ge20$ K.
Our neutron scattering experiments were carried out on the HB-1 and BT-9 thermal triple-axis spectrometers at the
high-flux isotope reactor, Oak Ridge National Laboratory and NIST Center for Neutron Research respectively.
We define the wave vector $Q$ at ($q_x,q_y,q_z$) as ($H,K,L$) = ($q_xa/2\pi,q_ya/2\pi,q_zc/2\pi$) reciprocal lattice unit (r.l.u.) in the tetragonal unit cell of PLCCO (space group $I4/mmm$, $a = b = 3.98$ \AA, and $c = 12.28$ \AA).
For the specific heat measurements, we used a commercial calibrated calorimeter with a 14 T physical property measurement system and a $^3$He insert capable of reaching 0.4 K.

\begin{figure}
\includegraphics[scale=.55]{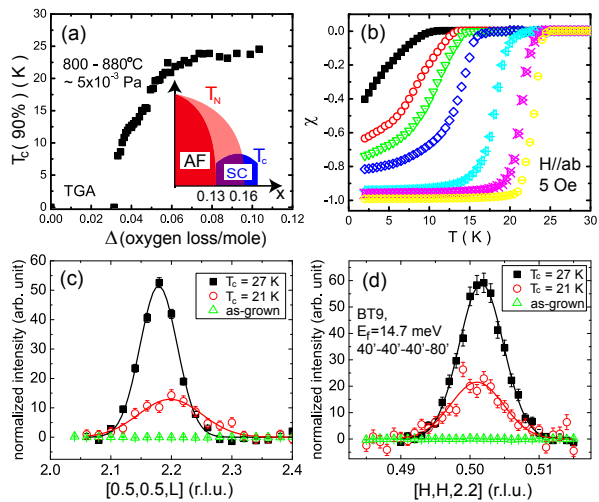}
\caption{(Color online) (a) $T_c$ (90\%) vs. $\Delta$ for the {\it same} sample annealed at various temperatures in vacuum. $T_c$ (90\%) is defined as the temperature where the susceptibility lose 90\% of its value at 2 K. The inset shows the schematic Ce content dependence of $T_N$ and $T_c$. (b) Magnetic susceptibility of PLCCO samples with $\delta$ = 0.034, 0.038, 0.042, 0.047, 0.053, 0.066, 0.092 from left to right of figure. (c) and (d) are the $[0.5,0.5,L]$ and $[H,H,2.2]$ scans through the impurity phase R$_2$O$_3$ peak position.  The scattering intensity is normalized to a
 measured phonon at $Q=(-0.1,-0.1,6)$ and $T = 80$ K \cite{wilson3}. 
}
\end{figure}

Recently, we suggested that the annealing process in electron-doped cuprates mainly
 involves repairing Cu deficiencies in the as-grown sample \cite{kang}, which suppresses local
superconductivity and induces staggered AF moments extending over several unit cells much
like the effect of nonmagnetic Zn impurities \cite{tarascon,sugiyama}. As a consequence,
the band filling and band parameters are not much affected by the annealing process \cite{gauthier,richard}. Assuming that the annealing process indeed fills the copper vacancies by producing thin slabs of the anion-deficient fluorite impurity phase R$_2$O$_3$ (R=Pr,La,Ce) through R$_2$Cu$_{1-f}$O$_{4-\delta}$ = (1-$f$)(R$_2$CuO$_{4-\beta}$)+$f$(R$_2$O$_3$)+[$-\delta$+$\beta$+$f(1-\beta$)](O) where
oxygen loss $\Delta=-\delta+\beta+f(1-\beta)$ \cite{kang}, the amount of the impurity R$_2$O$_3$ created during the annealing process should increase with increasing $\Delta$. Figures 1(c) and 1(d) show the $[0.5,0.5,L]$ and $[H,H,2.2]$ scans through the R$_2$O$_3$ impurity peak position at $(0.5,0.5,2.2)$ respectively \cite{matsuura,mang}. These data reveal that the impurity peaks for the higher $T_c$ samples have a larger peak intensity and narrower peak width along the $L$-axis direction. The volume fraction of the impurity phase increases with increasing $\Delta$.
Quantitatively, we estimate that the ratio of impurity contents between the $T_c=27$ K and 21 K samples
is about 1.6, meaning that the $T_c=21$ K sample should have $\sim$0.75\% Cu vacancies if we assume that the $T_c=27$ K
PLCCO has no Cu deficiency \cite{kang}.

\begin{figure}
\includegraphics[scale=.45]{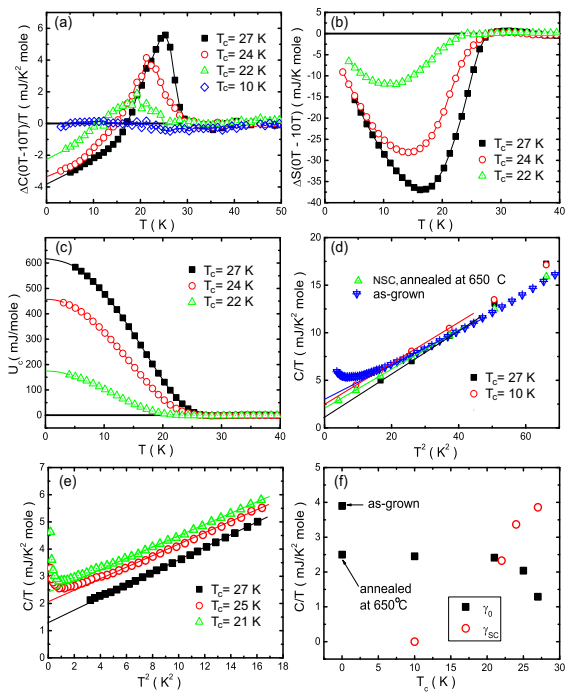}
\caption{(Color online)(a) The subtracted electronic specific heat for four different samples with field $H||c$. A small linear background ($<$ 0.35 mJ/K$^2$ mole at 30 K) is also subtracted for all the samples. (b) The entropy difference between the superconducting and normal states of three different $T_c$ PLCCO. (c) The temperature
dependence of $U_c$ for three different samples. (d) Temperature dependence of
$C/T$. (e) $C/T$ {\it vs.} $T^2$ measured on He$^3$ probe. (f) The residual electronic specific heat $C = \gamma_{0}T$ at 0 K and the superconducting electronic specific heat $C = \gamma_{SC}T$ at 0 K.
}
\end{figure}

Although Figures 1(a) and 1(b) show that the $T_c$ of PLCCO measured by susceptibility
increases with increasing $\Delta$, it is unclear when bulk superconductivity first appears in the annealing process.
Electronic specific heat measurements are one way to determine bulk superconductivity.
Since the upper critical field $H_{c2}$ of electron-doped superconductors is less than 10 T for $H||c$, the change of the electronic specific heat can be obtained from the difference between specific heats in
the superconducting ($H = 0$ T) and field-suppressed normal states ($H = 10$ T) \cite{balci}.
Figure 2(a) shows that the superconducting jump decreases rapidly with decreasing $T_c$ and vanishes for $T_c < 20$ K samples, although the superconducting volume determined by magnetic susceptibility is still large. Figures 2(b) and 2(c) plot the differential entropy $\Delta S$ and the superconducting condensation energy $U_c$ using
$\Delta S(T) = \int^{T}_{0}\left(\frac{C_{SC}-C_{n}}{T'}\right)dT'$,$
U_c(T) = \int^{T}_{T_c}\left(\Delta S(T')\right)dT'$ \cite{wilson3}, which further illustrates how fast the bulk superconductivity disappears with decreasing $T_c$.

To extract the absolute values of the electronic specific heat, we fit the low
temperature total specific heat data as
$C_{total}/T = \gamma_0  + \beta T^2$, where $\beta T^3$ is the lattice contribution and $\gamma_0T$ represents the electronic contribution. For as-grown PLCCO, the large low-temperature Schottky anomaly prevents an accurate determination of $\gamma_0$.  For annealed nonsuperconducting and superconducting samples, we obtain $\gamma_0$ using data in Fig. 2(d) and 2(e) and a measure of the residual superconducting contribution, $\gamma_{SC}$, using data in Fig. 2(a) with $\Delta C/T$(10 T$-$0 T). The $T_c$ dependence of $\gamma_0$ and $\gamma_{SC}$ are shown in Fig. 2(f).
The non-zero $\gamma_0$ may arise from the residual metallic nonsuperconducting portion of the superconductor \cite{balci}. Previous experiments have shown the existence of static 3D AF order coexisting with superconductivity in lower $T_c$ samples \cite{wilson2}. If a low-$T_c$ PLCCO sample is electronically phase-separated into insulating AF ordered, metallic nonsuperconducting, and superconducting regions \cite{wilson2}, one would expect that the residual $\gamma_0$ associated with the nonsuperconducting part of the metallic phase decreases with increasing $U_c(0)$ and volume fraction. In this picture, the annealing process initially weakens the static AF order with almost no change in the metallic phase, as indicated by the reduction in $\gamma_0$ from the as-grown to 650$^\circ$C annealed nonsuperconducting states [Fig. 2(f)]. Bulk superconductivity only emerges for PLCCO with $T_c\ge$20 K.
With further annealing, $\gamma_{SC}$ increases rapidly with increasing $T_c$ and $U_c$ while $\gamma_0$ decreases slowly, suggesting that the enhanced superconductivity with annealing is at the expense of static AF order.
For PLCCO samples with $T_c\ge 24$ K, where the static AF order essentially vanishes \cite{wilson},
the increasing $U_c$ with annealing may be at the
expense of the metallic nonsuperconducting phase with a reduction in $\gamma_0$ [Fig. 2(f)].

\begin{figure}
\includegraphics[scale=0.6]{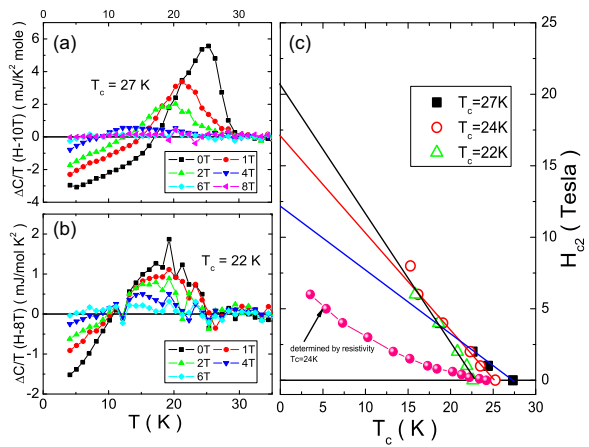}
\caption{(Color online)
Magnetic field dependence of the superconducting heat capacity anomalies for (a) $T_c$ = 27 K
and; (b) 22 K PLCCO samples. (c) A comparison of the upper critical field $H_{c2}$ determined by specific heat measurements for various $T_c$ PLCCO samples. $C(T)$ for $T_c$ = 24 K sample has already been published elsewhere \cite{wilson3}. For comparison, the $H_{c2}$ of the $T_c$ = 24 K sample obtained from resistivity is also shown. Solid linear lines are guided by eye. 
}
\end{figure}

Figures 3(a) and 3(b) show the magnetic field dependence of the superconducting specific heat anomaly for the $T_c = 27$ and 22 K samples respectively. Figure 3(c) plots the temperature dependence of $H_{c2}$ determined by the middle point of the negative slope region in the specific heat jump. Using the Helfand-Werthamer formula $H_{c2}(0) \approx 0.7 \times T_c \times dH_{c2}/dT$ \cite{helfand}, we find the $H_{c2}(0)$ values of 8.5 T, 12 T and 15 T for $T_c$ = 27 K, 24 K and 22 K respectively \cite{fournier,balci2,yayu}. For pure superconducting $T_c$ = 27 K PLCCO, the in-plane coherence length $\xi(0)$ is about 62 {\AA} in the clean $s$-wave limit \cite{fournier,shan}, where $H_{c2}(0)=\phi_0/2\pi\xi^2(0)$ and $\phi_0$ is the quantum of flux.
For $T_c$ = 24 K and 22 K samples, $\xi(0)$ decreases to 52 {\AA} and 47 {\AA}, respectively.
This is consistent with the expected 0.75\% Cu vacancies in the $T_c$ = 21 K sample,
where the average distance between two Cu vacancies is about 50 {\AA}.  On the other hand, dynamic spin-spin correlations at $\hbar\omega=1.5$ meV decreases with increasing $T_c$, changing from $200\pm21$ \AA\ to
$93\pm 14$ \AA\ and $80\pm10$ \AA\ for $T_c=21$ K and $T_N=40$ K, $T_c=23$ K and $T_N=25$ K, and $T_c=24$ K PLCCO  respectively \cite{wilson2}. 
Therefore, oxygen annealing should affect spin correlations determined by two-axis measurements \cite{motoyama}.

To test if the annealing process also affects the recently discovered magnetic resonance \cite{wilson}, we carried out neutron scattering experiments on $T_c$ = 27 K and $T_c$ = 21 K samples. Similar to measurements on $T_c$ = 24 K samples \cite{wilson},
we isolate the resonance ($E_r$) by taking the difference of constant $Q$-scans at $Q=(0.5,0.5,0)$ between the superconducting ($T = 6$ K) and normal states ($T = 30$ K). The data of constant $Q$-scans for the $T_c=27$ K and $T_c=21$ K sample are shown in Fig. 4(a) and 4(b) respectively. By fitting the subtracted data with Gaussians, we find $E_r = 10.3\pm0.38$ meV [Fig. 4(c)] and $E_r = 9.3\pm0.69$ meV [Fig. 4(d)].
The decreasing energy $E_r$ of the resonance with decreasing $T_c$ is consistent with the reduction of the  superconducting gap \cite{shan} and the $E_r=5.8k_BT_c$ universal plot \cite{wilson}. Figures 4(e) and 4(f) show constant energy scans near $E_r$ at 6 K and 30 K for these two samples.  Comparison of these data also reveals that the intensity gain from the normal to superconducting states for the $T_c = 27$ K sample is considerably larger than that of the $T_c = 21$ K sample in agreement with our conclusions regarding bulk superconductivity based on thermodynamic measurements.

\begin{figure}
\includegraphics[scale=.45]{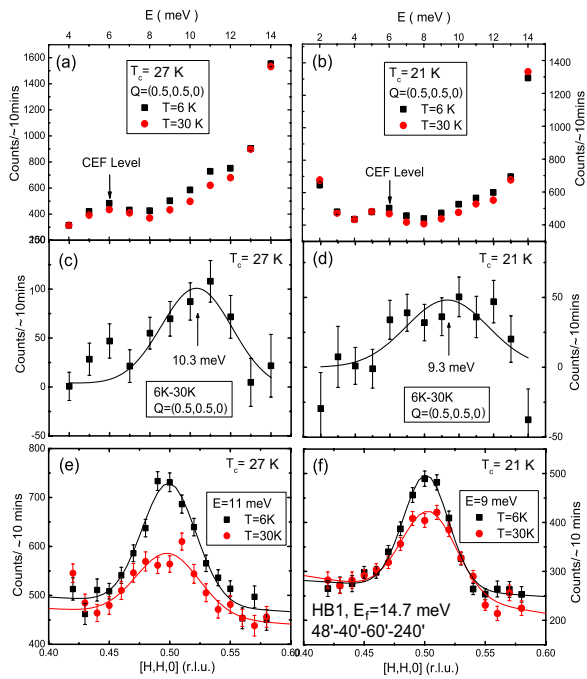}
\caption{(Color online) Constant ${Q}$ scans at $Q=(0.5,0.5,0)$ at $T = 6$ K and 30 K for (a) $T_c = 27$ K and (b) $T_c = 21$ K PLCCO. The small peak at $\hbar\omega=6$ meV is the Pr crystal field from the ``impurity'' R$_2$O$_3$ phase which is not present in the as-grown PLCCO \cite{wilson}. The corresponding differences between 6 K and 30 K are shown in (c) and (d), indicating resonance peaks. (e) and (f) show constant energy scans through
$[H,H,0]$ at 6 K and 30 K near the resonance energies for these two samples.
}
\end{figure}

\section{DISCUSSIONS and CONCLUSIONS}
The effect of annealing on superconductivity and magnetism in PLCCO
appears to be similar to that of Zn-doping in the electron-doped superconductor
NCCO, where a few percent of Zn impurities (Cu vacancies in
the as-grown sample) can completely suppress superconductivity \cite{tarascon,sugiyama}.
In addition, the increasing $\gamma_0$ with decreasing $T_c$ in PLCCO [Fig. 2(f)] is also similar to the Zn-doping dependence of $\gamma_0$ in
YBCO \cite{sisson}. Since both the superconducting coherence length and spin-spin correlation length are strongly affected by the oxygen annealing process, 
these results call into question the suggestion that there is no genuine
coexisting AF and superconducting phase and the QCP occurs just before the superconducting dome at $x\approx 0.13$ \cite{motoyama}. 
In our view, any attempt to establish the Ce-doping evolution of the AF to superconducting phase transition on electron-doped materials must begin by systematically determining the oxygen annealing effect at different Ce-dopings \cite{tanaka}.
This implies that
one must examine the electron-doped superconductors with the highest $T_c$ at a fixed Ce-doping
as in the thin film PCCO case
to determine their intrinsic electronic properties \cite{dagan,yu}.

\begin{acknowledgments}
The neutron scattering work was supported in part by the
U.S. NSF DMR-0453804. The PLCCO single crystal growth at UT was
supported by the U.S. DOE BES under grant No. DE-FG02-05ER46202.
ORNL was supported by U.S. DOE contract
DE-AC-05-00OR22725 through UT/Battell, LLC.
Work at IOP was supported by CAS ITSNEM projects 2006CB601000 and
2006CB92180.
\end{acknowledgments}



\begin{thebibliography}{}
\bibitem{kastner} M. A. Kastner, R. J. Birgeneau, G. Shirane, and Y. Endoh, Rev. Mod. Phys. {\bf 70}, 897 (1998). 
\bibitem{wakimoto} S. Wakimoto, G. Shirane, Y. Endoh, K. Hirota, S. Ueki, K. Yamada, R. J. Birgeneau,
M. A. Kastner, Y. S. Lee, P. M. Gehring, and S. H. Lee, Phys. Rev. B {\bf 60}, R769 (1999).
\bibitem{uefuji} T. Uefuji, K. Kurahashi, M. Fujita, M. Matsuda, and K. Yamada, Physica C {\bf 378-381},273 (2002).
\bibitem{fujita} M. Fujita, T. Kubo, S. Kuroshima, T. Uefuji, K. Kawashima, and K. Yamada, Phys. Rev. B {\bf 67}, 014514 (2003).
\bibitem{dagan} Y. Dagan, M. M. Qazilbash, C. P. Hill, V. N. Kulkarni,
and R. L. Greene, Phys. Rev. Lett. {\bf 92}, 167001 (2004).
\bibitem{yu} W. Yu, J. S. Higgins, P. Bach, and R. L. Greene, Phys. Rev. B {\bf 76}, 020503(R) (2007).
\bibitem{zimmer} A. Zimmers, J. M. Tomczak, R. P. S. M. Lobo, N. Bontemps, C. P. Hill,
M. C. Barr, Y. Dagan, R. L. Greene, A. J. Millis, and C. C. Homes, Europhys. Lett. {\bf 70}, 225 (2005).
\bibitem{matsui} H. Matsui, T. Takahashi, T. Sato, K. Terashima, H. Ding, T. Uefuji,
and K. Yamada, Phys. Rev. B {\bf 75}, 224514 (2007).
\bibitem{onose} Y. Onose, Y. Taguchi, K. Ishizaka, and Y. Tokura, Phys. Rev. B {\bf 69}, 024504 (2004).
\bibitem{motoyama} E. M. Motoyama, G. Yu, I. M. Vishik, O. P. Vajk, P. K. Mang, and M. Greven, Nature (London) {\bf 445}, 186 (2007).
\bibitem{tokura} Y. Tokura, H. Takagi, and S. Uchida, Nature (London) {\bf 337}, 345 (1989).
\bibitem{takagi} H. Takagi, S. Uchida, and Y. Tokura, Phys. Rev. Lett. {\bf 62}, 1197 (1989).
\bibitem{higgins} J. S. Higgins, Y. Dagan, M. C. Barr, B. D. Weaver, and R. L. Greene, Phys. Rev. B {\bf 73}, 104510 (2006).
\bibitem{gauthier} J. Gauthier, G. Gagn$\rm \acute{e}$, J. Renaud, M.-$\rm \grave{E}$. Gosselin,
and P. Fournier, Phys. Rev. B {\bf 75}, 024424 (2007).
\bibitem{kang} H. J. Kang, Pengcheng Dai, B. J. Campbell, P. J. Chupas, S. Rosenkranz, P. L. Lee,
Q. Huang, S. L. Li, S. Komiya, and Y. Ando, Nature Materials {\bf 6}, 224 (2007).
\bibitem{kim} J. S. Kim and D. R. Gaskell, Physica C {\bf 209}, 381 (1993).
\bibitem{jiang} W. Jiang, S. N. Mao, X. X. Xi, X. Jiang, J. L. Peng, T. Venkatesan, C. J. Lobb,
and R. L. Greene, Phys. Rev. Lett. {\bf 73}, 1291 (1994).
\bibitem{wilson2} Pengcheng Dai, H. J. Kang, H. A .Mook, M. Matsuura, J. W. Lynn, Y. Kurita,
S. Komiya, and Y. Ando Phys. Rev. B {\bf 71}, 100502(R) (2005);
S. D. Wilson, S. L. Li, Pengcheng Dai, W. Bao, J. H. Chung, H. J. Kang, S. H. Lee, S. Komiya,
Y. Ando, and Q. Si, {\bf 74}, 144514 (2006).
\bibitem{wilson} S. D. Wilson, Pengcheng Dai, S. L. Li, 
S. X. Chi, H. J. Kang, and J. W. Lynn, Nature (London) {\bf 442}, 59 (2006).
\bibitem{tarascon} J. M. Tarascon, E. Wang, S. Kivelson, B. G. Bagley, G. W. Hull, and
R. Ramesh, Phys. Rev. B {\bf 42}, 218 (1990).
\bibitem{sugiyama} J. Sugiyama, S. Tokuono, S. Koriyama, H. Yamauchi, and S. Tanaka, 
Phys. Rev. B {\bf 43} 10489 (1991).
\bibitem{richard} P. Richard, M. Neupane, Y.-M. Xu, P. Fournier, S. Li, Pengcheng Dai, Z. Wang, and H. Ding, Phys. Rev. Lett. {\bf 99}, 157002 (2007).
\bibitem{matsuura} M. Matsuura, Pengcheng Dai, H. J. Kang, J. W. Lynn, D. N. Argyriou, K.
Prokes, Y. Onose, and Y. Tokura, Phys. Rev. B {\bf 68}, 144503 (2003).
\bibitem{mang} P. K. Mang, S. Larochelle, A. Mehta, O. P. Vajk, A. S. Erickson, L. Lu, W. J. L. Buyers, A.
F. Marshall, K. Prokes, and M. Greven, Phys. Rev. B {\bf 70}, 094507 (2004).
\bibitem{balci} H. Balci and R. L. Greene, Phys. Rev. B {\bf 70}, 140508 (2004).
\bibitem{wilson3} S. D. Wilson, S. L. Li, J. Zhao, G. Mu, H.-H. Wen, J. W. Lynn, P. G. Freeman,
L. P. Regnault, K. Habicht, and Pengcheng Dai, PNAS {\bf 104}, 15259 (2007).
\bibitem{helfand} E. Helfand and N. R. Werthamer, Phys. Rev. {\bf 147}, 288 (1966).
\bibitem{fournier} P. Fournier and R. L. Greene, Phys. Rev. B {\bf 68}, 094507 (2003).
\bibitem{balci2} H. Balci, C. P. Hill, M. M. Qazilbash, and R. L. Greene, Phys. Rev. B {\bf 68}, 054520 (2003).
\bibitem{yayu} Y. Wang, S. Ono, Y. Onose, G. Gu, Y. Ando, Y. Tokura, S. Uchida, and N. P. Ong, Science {\bf 299}, 86 (2003).
\bibitem{shan} L. Shan, Y. Huang, Y. L. Wang, S. L. Li, J. Zhao, Pengcheng Dai, Y. Z. Zhang, C. Ren, and H. H. Wen, Phys. Rev. B {\bf 77}, 014526 (2008).
\bibitem{sisson} D. L. Sisson, S. G. Doettinger, A. Kapitulnik, R. Liang,
D. A. Bonn, and W. N. Hardy, Phys. Rev. B {\bf 61}, 3604 (2000).
\bibitem{tanaka} Y. Tanaka, T. Motohashi, M. Karppinen, and H. Yamauchi, J. Solid State Chemistry {\bf 181}, 365 (2008).
\end{thebibliography}
\end{document}